\documentclass[prc,twocolumn,showpacs,floatfix,superscriptaddress]{revtex4-1}
\usepackage{hyperref}
\usepackage{graphicx}
\usepackage{amssymb}
\usepackage{amsmath}
\usepackage{mathrsfs}
\usepackage{bm}
\usepackage{enumerate}
\usepackage{multirow}
\usepackage{hhline}
\usepackage{color}

\begin{document}

\title{Charged Ising Model of Neutron Star Matter}

\author{K.H.O.~Hasnaoui}
\affiliation{
Department of Physics, Florida State University, Tallahassee, Florida 32306, USA
}
\author{J.~Piekarewicz}
\affiliation{
Department of Physics, Florida State University, Tallahassee, Florida 32306, USA
}

\date{\today}
\begin{abstract}
\begin{description}
 \item[Background] The inner crust of a neutron star is believed to consist 
 of Coulomb-frustrated complex structures known as {\sl ``nuclear pasta''} 
 that display interesting and unique low-energy dynamics. 
 \item[Purpose] To elucidate the structure and composition of the neutron-star
 crust as a function of temperature, density, and proton fraction.
 \item[Methods] A new lattice-gas model, the {\sl ``Charged-Ising 
 Model"} (CIM), is introduced to simulate the behavior of neutron-star matter.
 Preliminary Monte Carlo simulations on $30^3$ lattices are performed for a variety 
 of temperatures, densities, and proton fractions.
 \item[Results] Results are obtained for the heat capacity, pair-correlation 
 function, and static structure factor for a variety of conditions appropriate
 to the inner stellar crust. 
 \item[Conclusions]  Although relatively simple, the CIM captures the essence 
 of Coulomb frustration that is required to simulate the subtle dynamics of the 
 inner stellar crust. Moreover, the computationally demanding long-range 
 Coulomb interactions have been pre-computed at the appropriate lattice sites 
 prior to the start of the simulation resulting in enormous computational gains. 
 This work demonstrates the feasibility of future CIM simulations 
 involving a large number of particles as a function of density, temperature, 
 and proton fraction.
\end{description}
\end{abstract}


\maketitle

\section{Introduction}
\label{Introduction}
Neutron stars are compact objects with radii of the order of ten
kilometers and masses comparable to that of the Sun. The solution 
of the Tolman-Oppenheimer-Volkoff (TOV)
equations\,\cite{oppenheimer1,bombaci1}, which prescribes the
structure of spherically symmetric, self-gravitating compacts object
in hydrostatic equilibrium, provides information on the density
profile of the star. Remarkably, the structure of neutron stars 
depends exclusively on the nuclear equation of state (EOS). 
Given the constraint of hydrostatic equilibrium, the density profile 
spans an enormous range of densities: from the extremely dilute 
crustal densities up to core densities that may greatly exceed 
nuclear-matter saturation density. Understanding what novels 
phases of matter emerge under these extreme conditions is both 
fascinating and unknown\,\cite{glendenning1,glendenning2}. Moreover, 
it represents one of the grand challenges in nuclear physics: 
{\sl``How does subatomic matter organize itself?''}\,\cite{NucPhys2012}.

The highest density attained in the stellar core depends critically on
the equation of state of neutron-rich matter.  Although at such high
densities the EOS is poorly constrained, it has been speculated that
many exotic phases may emerge under such extreme conditions. 
These may include pion or kaon condensates\,\cite{Ellis:1995kz,
Pons:2000xf}, strange quark matter\,\cite{Weber:2004kj}, and color
superconductors\,\cite{Alford:1998mk, Alford:2007xm}. It is also
often assumed that the uniform core may have a non-exotic component 
consisting of neutrons, protons, electrons, and muons in chemical 
equilibrium. However, at densities of about half of nuclear-matter
saturation density, the uniform core becomes unstable against cluster
formation. At these ``low'' densities the average inter-nucleon
separation increases to such an extent that it becomes energetically
favorable for the system to segregate into regions of normal density
(nuclear clusters) and regions of low density (neutron vapor). The
transition region between the homogeneous and non-homogeneous 
phases constitutes the crust-core interface. It is the aim of this work 
to study the structure and composition of the crust-core interface 
where distance scales are such that the Coulomb and nuclear interactions 
become comparable in strength. Under these unique conditions 
neutron-rich matter becomes {\sl ``frustrated"}. Frustration, a 
prevalent phenomenon characterized by the existence of a very large 
number of low-energy configurations, emerges from the impossibility
to simultaneously minimize all elementary interactions in the system. 
In the inner stellar crust this leads to a myriad of complex 
structures---collectively known as {\sl ``nuclear pasta''}---that are 
radically different in topology yet extremely close in energy. Moreover, 
due to the preponderance of low-energy states, frustrated systems 
display an interesting and unique low-energy dynamics.  For example, 
it has been speculated that pasta formation could enhance the coherent 
scattering of neutrinos from such exotic structures. This could have
important consequences on the supernova explosion mechanism and 
subsequent cooling dynamics\,\cite{watanabe1,horowitz1,margueron1}.

In this contribution we are interested in the equation of state of
neutron-rich matter at densities of relevance to the inner stellar
crust \cite{chamel1,bertulani1}. We will model this charge-neutral
system in terms of its basic constituents, namely, neutrons, protons,
and an ultra-relativistic Fermi gas of electrons. In particular, no ad-hoc
biases will be introduced in regard to the structure of the exotic pasta shapes 
({\sl i.e.,} whether they form droplets, rods, slabs, bubbles, {\sl etc.}).
Rather, we will allow the clustering to develop dynamically from an
initial (random) configurations of nucleons. The aim of this work is
to explore the dynamics of the system as a function of density,
temperature, and proton fraction. Note that the original work by
Ravenhall and collaborators was carried out at zero temperature
in a mean-field approach\,\cite{ravenhall1,lorentz1,ravenhall2}.
Recently, more sophisticated approaches---based on Monte Carlo 
and Molecular Dynamics simulations\,\cite{horowitz1,horowitz2,
horowitz3,maruyama1,watanabe1,watanabe2,watanabe3,watanabe4,
watanabe5,watanabe6,watanabe7,watanabe8,watanabe9}, a 
Dynamical-Wavelet approach\,\cite{sebille1,sebille2,sebille3}, 
relativistic mean-field calculations\,\cite{maruyama2,maruyama3,
avancini1, avancini2,grill1}, and Skyrme-Hartree-Fock 
methods\,\cite{newton}---have been implemented and have 
confirmed the existence of these exotic phases at very low 
temperatures and moderate proton fractions. However, given that 
chemical equilibrium suggests that the proton fraction in the inner 
stellar crust is very low---indeed, significantly lower than
normally assumed---it has recently been put into question 
whether pasta formation is even possible in such proton-poor
environments\,\cite{piekarewicz1}.  Moreover, simulations at different
temperatures are both critical and interesting because the long-range
Coulomb interaction is responsible for the extreme fragility of
crystals. That is, the melting (or charge-ordering) temperature in
crystals $T_{c}$ is significantly smaller than the relevant
Coulomb energy scale $E_{\rm Coul}\!=\!e^{2}/a$ (here $a$ is the 
lattice spacing).  Such an energy mismatch
introduces a large temperature gap
($k_{B}T_{c}\!<\!k_{B}T\!\ll\!E_{\rm Coul})$ where the system displays
unconventional pasta-like behavior that reflects the strong
frustration induced by the long-range interactions. In particular,
condensed-matter simulations with long-range interactions have
reported the opening of a {\sl pseudogap} in the density of states 
in response to the strong frustration\,\cite{Pramudya2011}. This
unconventional pseudogap region mediates the transition from 
the Wigner Crystal to the Fermi liquid. Interestingly enough, the 
pseudogap disappears for a system with only short-range 
interactions.

As an alternative to the numerically intensive Monte Carlo and 
Molecular Dynamics simulations, we introduce here the 
{\sl ``Charged Ising Model''} (CIM). The CIM is a {\sl lattice-gas} 
model that while simple in its assumptions, retains the essence 
of Coulomb frustration. Numerical simulations based on this model 
are not as computationally demanding because the long-range 
Coulomb interaction, computed here exactly via an Ewald 
summation, may be pre-computed at the appropriate lattice
sites and then stored in memory prior to the start of the simulation.
This represents an enormous advantage when trying to simulate 
systems with a large number of particles as a function of temperature, 
density, and proton fraction. It is an important goal of this work to
extend earlier (fixed-temperature) approaches by studying the 
thermal properties of the crust. Specifically, we rely on classical 
Monte Carlo simulations of the CIM to investigate phase transitions 
in stellar matter in the presence of Coulomb frustration. The CIM is 
reminiscent of an earlier approach 
developed in Refs.\,\cite{ising_star1,ising_star2,ising_star3}. Yet, it
improves on it in two respects: (a) by including explicitly the isospin 
degree of freedom that is required for a proper treatment of asymmetric 
matter and (b) by using the Ewald summation to properly treat the 
long-range Coulomb interaction. Although limited in their treatment 
of quantum fluctuations, classical simulations like the ones proposed 
here are essential to uncover correlations that go beyond mean-field 
approaches. In particular, both spatial and thermal correlations---as 
embodied in the static structure factor and heat capacity---will be 
computed as a function of density, temperature, and proton fraction 
in the search of signatures of phase transitions.

The paper has been organized as follow. In Sec.\,\ref{The model} the
CIM and the general framework will be introduced. Results from the
simulations will be presented in Sec.\,\ref{Results} and then compared
against previous findings reported in Ref.\,\cite{piekarewicz1}. Moreover,
we will extend this earlier work by following the evolution of the pasta 
structures as a function of the temperature at fixed density and proton
fraction. Finally, we offer our conclusions and suggestions for future 
work in Sec.\,\ref{Conclusions}.

\section{The CIM model: General framework}
\label{The model}

The main constituents of the stellar crust are neutrons, protons, and a 
background gas of neutralizing electrons. At the densities of relevance 
to the inner crust, the electrons may be treated as as an ultra-relativistic
Fermi gas, namely, with a dispersion relation $\epsilon(p)\!=\!p$. Although
the CIM presented here represents a simplification of the model first 
introduced in Ref.\,\cite{horowitz1}, it still retains the essence of Coulomb 
frustration, namely, competing interactions consisting of a short-range
nuclear interaction and a long-range Coulomb potential. The CIM assumes
that nucleons are allowed to occupy only the discrete sites of a 
three-dimensional cubic lattice of volume $V\!=\!L^{3}\!$ containing a total 
number of $S$ sites. The electrons on the other hand are assumed to 
provide a uniform neutralizing background. 

The potential energy consists of a sum of a short-range interaction between
nucleons and a long-range Coulomb interaction between protons and the
uniform electron background. That is,
\begin{equation}
 V_{\rm Total} = V_{\rm Nuclear} + V_{\rm Coulomb} \;. 
 \label{VTotal}
\end{equation}
For the short-range nuclear interaction the potential energy is assumed 
to be given by a sum of two-body terms that act exclusively over nearest 
neighbors. That is,
\begin{equation}
 V_{\rm Nuclear} = \frac{1}{2}\sum_{\langle i,j\rangle}^{S} v_{ij}n_{i}n_{j} \,, 
 \label{VNuclear}
\end{equation}
where $n_{i}\!=\!0,1$ denotes the occupation number of site $i$ and the
``elementary" two-body interaction is given by  
\begin{equation}
 v_{ij} = \Big(b+c\tau_{i}\tau_{j}\Big) \,.
 \label{VNTwoBody} 
\end{equation}
Here $\tau_{i}$ is the isospin of the nucleon occupying site $i$, with 
$\tau_{i}\!=\!+1$ for protons and $\tau_{i}\!=\!-1$ for neutrons. Note
that the repulsive short-range nature of the NN interaction is simulated
here by precluding the double occupancy of lattice sites. Also note that 
the two-body interaction is assumed to be isospin dependent to simulate 
quantum statistics. For example, in order to prevent pure neutron matter 
to be bound, the neutron-neutron interaction has to be made repulsive,
namely, $v_{nn}\!=\!(b+c)\!>\!0$. Indeed, we now describe the procedure
employed to fixed the two parameters $b$ and $c$. We assume that a
completely filled lattice containing $A\!=\!S$ nucleons corresponds to 
nuclear-matter at saturation density. That is,
\begin{equation}
 \rho_{{}_{0}} = \frac{A}{V} = \frac{1}{a^{3}} = 0.16\,{\rm fm}^{-3}\,
 \implies a=1.842\,{\rm fm}. 
 \label{aLatt}
\end{equation}
For such a filled lattice the energy per nucleon of symmetric nuclear matter
and pure neutron matter at saturation density are given by
\begin{subequations}
 \begin{align}
   & \frac{E_{\rm SNM}}{A} = 3(b-c) =  -16.5\,{\rm MeV} \,,\\
   & \frac{E_{\rm PNM}}{A} = 3(b+c) = +15.5\,{\rm MeV} \,.
 \end{align}
 \label{ENucMatt}
\end{subequations}
This choice fixes the two model parameters to the following values:
\begin{subequations}
 \begin{align}
 & b = -\frac{1}{6}\,{\rm MeV}  = -0.167\,{\rm MeV} \,, \\
 & c = +\frac{16}{3}\,{\rm MeV} =  5.333\,{\rm MeV} \,,
 \end{align}
 \label{bANDc}
\end{subequations}
or equivalently,
\begin{subequations}
 \begin{align}
 & v_{pn} = v_{np} = -5.500\,{\rm MeV} \,, \\
 & v_{pp} = v_{nn} = +5.167\,{\rm MeV} \,.
 \end{align}
 \label{vij}
\end{subequations}

\begin{figure}[htbp]
\begin{center}
 \includegraphics[width=\columnwidth]{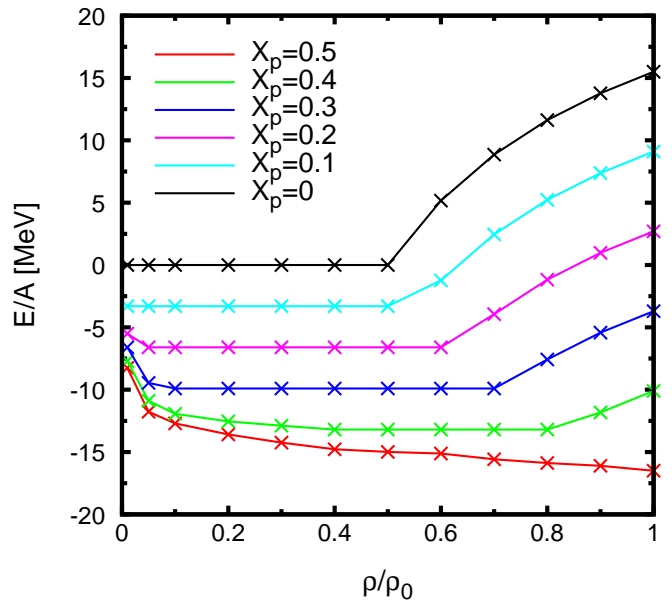}
 \end{center}
 \vspace{-0.4cm}
\caption{(color online) Energy per nucleon as a function of filling fraction 
 $\rho/\rho_{{}_{0}}\!=\!A/S$ and proton fraction $x_{p}\!=\!Z/A$ for infinite 
 nuclear matter. A proton fraction of $x_{p}\!=\!0.5$ represents symmetric 
 nuclear matter whereas $x_{p}\!=\!0$ corresponds to pure neutron matter. 
 Monte Carlo simulations were performed on a lattice with $S\!=\!(20)^{3}$ 
 sites as the temperature $T\!\rightarrow\!0$.}
\label{fig:EOSneutral}
\end{figure}

To illustrate the dynamics behind this very simple choice we display in 
Fig.\ref{fig:EOSneutral} results for the energy per nucleon of infinite 
nuclear matter (with the Coulomb interaction turned off and no electrons) 
as a function of both the filling fraction $\rho/\rho_{{}_{0}}\!=\!A/S$ and the 
proton fraction $x_{p}\!=\!Z/A$. Monte-Carlo simulations were performed 
on a cubic lattice of $S\!=\!(20)^{3}$ sites and at a temperature of 
$T\!\approx\!0$. Note that all simulations were started at the high 
temperature of $T\!=\!20\,{\rm MeV}$ and slowly cooled down to 
$T\!\approx\!0$ until the configuration was frozen. The EOS
for symmetric nuclear matter ($x_{p}\!=\!0.5$) yields, by construction, a binding 
energy per nucleon at saturation density of $16.5\,{\rm MeV}$ and decreases to 
about $8\,{\rm MeV}$ at very low densities---corresponding to the binding energy 
of an isolated (symmetric) cluster. Note that in contrast to mean-field descriptions 
that assume nuclear matter to be uniform---and thus the energy to vanish at very 
low densities---the lattice-gas model takes full account (at least classically) of 
clustering correlations.
At the other extreme ($x_{p}\!=\!0$) pure neutron matter is unbound at all 
densities and yields, by construction, an energy per neutron at saturation 
density of $15.5\,{\rm MeV}$; this corresponds to a symmetry energy at
saturation density of $32\,{\rm MeV}$. Note that in the lattice model the
energy of pure neutron matter vanishes at half filling (and below) as the
lowest energy configuration consists of neutrons surrounded by empty
sites.

For the emergence of frustration and the concomitant development of pasta 
structures, the Coulomb repulsion between protons is of critical importance. 
As mentioned earlier, at densities of relevance to the bottom layers of the
inner crust ({\sl i.e.,} $10^{13}$-$10^{14}\,{\rm g/cm}^{3}$) the competition 
between the short-range nuclear attraction and the long-range Coulomb 
repulsion is the main driving force behind frustration. Whereas in earlier 
publications we adopted an approximate {\sl screened} Coulomb 
interaction\,\cite{horowitz1}, more recently\,\cite{piekarewicz1} we have 
treated the problem exactly by means of an Ewald summation\,\cite{ewald1}.
We follow the exact Ewald treatment here as well. 

Using Ewald's method we can cast the Coulomb potential as a sum of two-body 
interactions plus a constant term. That is,
\begin{equation}
 V_{\rm Coulomb} = V_{0} + 
 \frac{v_{{}_{0}}}{2}\sum_{i\ne j}^{S} u_{ij}n_{i}^{p}n_{j}^{p} \,, 
 \label{VCoulomb}
\end{equation}
where $n_{i}^{p}\!\equiv\!n_{i}(1\!+\!\tau_{i})/2$ denotes the {\sl proton} occupation 
number of site $i$, $v_{{}_{0}}\!\equiv\!e^{2}/L$ sets the Coulomb energy scale, and 
$V_{0}$ is an overall constant\,\cite{piekarewicz1}. The dimensionless two-body 
potential $u_{ij}$ may be written in terms of short- and long-range contributions: 
\begin{align}
  u_{ij} &= \Big[u_{\rm sr}({\bf s}_{ij}) +
                    u_{\rm lr}({\bf s}_{ij}) \Big] =
                    \frac{{\rm erfc}(s_{ij}/s_{0})}{s_{ij}} \nonumber \\
                    &+ \sum_{{\bf l}\ne 0} 
                    \frac{\exp(-\pi^{2}s_{0}^{2}\,l^{2})}{\pi l^{2}} 
                    \exp(-2\pi i{\bf l}\cdot{\bf s}_{ij}) \;,
 \label{UCoulEwald}
\end{align}
where ${\bf l}\!=\!(l_{x},l_{y},l_{z})$ represents a triplet of integers and 
${\bf s}_{ij}$ is the separation between lattice sites $i$ and $j$ in 
dimensionless units. We now proceed with a brief explanation of the 
various terms; for a more detailed account see Ref.\,\cite{piekarewicz1}. 
The Coulomb potential is an interaction with no intrinsic scale. Ewald 
introduced a scale into the problem by adding $Z$ positive and $Z$ negative 
{\sl smeared} charges at the exact location of each proton. It is both customary 
and convenient to introduce a gaussian charge distribution with a smearing 
parameter $a_{s}$; in the above expression $s_{0}\!=\!a_{s}/L$. The 
role of each negative charge is to fully screen the corresponding point proton 
charge over distances of the order of the smearing parameter. Thus, as long
as $a_{s}$ is significantly smaller than the box length $L$, the resulting (screened) 
two-body potential [${\rm erfc}(s/\!s_{0})\!/\!s$] will become short ranged and thus 
amenable to be treated using the minimum-image convention\,\cite{ewald2,ewald3}.
What remains then is a periodic system of smeared positive charges together 
with the neutralizing electron background. Whereas in configuration space this
{\sl long-range} contribution is slowly convergent, the great merit of the Ewald 
construction is that it can be made to converge rapidly if evaluated in momentum
space, namely, as a Fourier sum. Indeed, the Fourier sum is rapidly convergent 
because (dimensionless) momenta {${\bf l}$ satisfying $ls_{0}\!\gg\!1$ make a 
negligible contribution to the Fourier sum. Hence, by suitably tuning the value of 
the smearing parameter, the evaluation of the Coulomb potential may be written 
in terms of two rapidly convergent sums; one in configuration space and one in 
momentum space\,\cite{piekarewicz1}. This is the enormous advantage of the
Ewald construction. Another enormous advantage---now specific to the lattice 
model---is that one may pre-compute the two-body Coulomb interaction $u_{ij}$ 
for all different pairs of lattice sites and then stored them in an array for later 
retrieval during the simulation. 

In what follows we employ a canonical ensemble to perform numerical 
simulations of a system consisting of $A$ nucleons, $Z\!=\!x_{p}A$, 
protons, and temperature $T$. A configuration in the system may be
specified by a collection of $S$ occupation numbers 
${\pmb\alpha}\!=\!(\alpha_{1},\alpha_{2},\ldots,\alpha_{S})$, where at 
each site $\alpha_{i}\!=\!\{p,n,0\}$, depending on whether the site is 
occupied by either a proton or a neutron, or it remains vacant. Given 
that the potential energy is independent of momentum, the partition 
function for the system factors into a product of a partition function 
in momentum space---that has no impact in the computation of 
momentum-independent observables---times a coordinate space 
(or interaction) partition function of the form:
\begin{equation}
 {\mathcal Z(A,x_{p},T)} = \sum_{\pmb\alpha}
 \exp\Big(\!-\!\beta\,V_{\rm Total}({\pmb\alpha})\Big) \;,
 \label{PartFcn}
 \end{equation}
where $\beta\!=\!(k_{B}T)^{-1}$ is the inverse temperature. In turn,
the expectation value of any momentum-dependent observable 
${\mathcal O}$ may be estimated by performing the appropriate
statistical average. That is,
\begin{equation}
  \langle{\mathcal O}\rangle = \sum_{\pmb\alpha} 
  {\mathcal O}({\pmb\alpha})P_{\pmb\alpha}(T)\;,
  \label{OAverage}
 \end{equation}
where $P_{\pmb\alpha}(T)$ represents the probability of finding
the system in a given configuration $\pmb\alpha$. In the canonical
ensemble such a probability is proportional to the properly normalized
Boltzmann factor:
\begin{equation}
 P_{\pmb\alpha}(T)=
 \frac{\exp\Big(\!-\!\beta\,V_{\rm Total}({\pmb\alpha})\Big)}
 {\mathcal Z(A,x_{p},T)} \;.
 \label{Palpha}
 \end{equation}
Given that the momentum-independent interactions have no impact on the
kinetic energy of the system, the expectation value of the kinetic energy
reduces to a sum of a classical contribution for the nucleons and a 
quantum contribution for the electrons. That is,
\begin{equation}
  \langle K \rangle = \frac{3}{2}Ak_{B}T + 
  \frac{3}{4}Zk_{\rm F}\left[1+
  \frac{2\pi^{2}}{3}\left(\frac{T}{T_{\rm F}}\right)^{2}\right]\;,
 \label{KEnergy}
\end{equation}
where $k_{\rm F}\!=\!k_{B}T_{\rm F}$ is the electronic Fermi momentum. The
total energy of the system is then given by
\begin{equation}
  \langle E(A,x_{p},T) \rangle = \langle K(A,x_{p},T) \rangle + 
  \langle{V_{\rm Total}(A,x_{p},T)}\rangle \;.
 \label{TotalEnergy}
\end{equation}
We note that the sum over ${\pmb\alpha}$ in Eq.\,(\ref{OAverage}) runs over a 
total number of configurations given by
\begin{equation}
  C(A,Z) = \frac{S!}{Z!(A-Z)!(S-A)!} \;.
 \label{Configurations}
\end{equation}
This number becomes astronomical even for systems of moderate size. Thus, 
to properly sample the statistical ensemble, we rely on a Metropolis Monte-Carlo 
algorithm \cite{metropolis1} to generate configurations distributed according to Eq.\,(\ref{Palpha}).
Given that the kinetic energy of the system corresponds to that a classical ideal 
gas of nucleons and an ultra-relativistic Fermi gas of electrons, their contribution
to the heat capacity is both known and smooth. Thus, any non-analytic behavior
associated with the existence of a phase transition must arise from the interactions.
For example, in the case of the heat capacity the potential energy contribution will
be estimated from the fluctuations in the potential energy. That is,
\begin{equation}
  \frac{C_{\rm v}}{k_{\rm B}}=\frac{3}{2}A + \pi^{2}Z\left(\frac{T}{T_{\rm F}}\right) +
  \frac{ \langle V_{\rm Total}^{2}\rangle - \langle V_{\rm Total}\rangle^{2}}
  {(k_{\rm B}T)^{2}} \;.
 \label{HeatCapacity}
\end{equation}
Whereas the heat capacity accounts for the mean-square energy fluctuations---which 
diverge near phase transitions---the static structure factor $S({\bf k})$ provides a 
complimentary observable associated with the mean-square {\sl density 
fluctuations}~\cite{fetter1}. Moreover, $S({\bf k})$ is intimately related to a quantity 
particularly suitable to be modeled in computer simulations, namely, the pair-correlation 
function $g({\bf r})$. Indeed, $S({\bf k})$ and $g({\bf r})$ are simply Fourier transforms of 
each other. The pair-correlation function $g({\bf r})$ is particularly simple to simulate as 
it represents the probability of finding a pair of particles separated by a fixed distance 
${\bf r}$. For a system containing $N$ particles and confined to a simulation volume 
$V$, $g(r)$ may be computed exclusively in terms of the instantaneous positions of the 
particles. That is,
\begin{equation}
   g({\bf r}) = 1 + \frac{V}{N(N-1)} 
   \Big\langle \sum_{i\ne j} \delta({\bf r}-{\bf r}_{ij})\Big\rangle \;,
\label{GofR}
\end{equation}
where ${\bf r}_{ij}\!=\!{\bf r}_{i}\!-\!{\bf r}_{j}$ and the ``{\sl brackets}'' represent an 
ensemble average. Note that g({\bf r}) is normalized to $1$ at very large distances. 
Whereas for a uniform fluid the one-body density is constant, interesting two-body 
correlations emerge as a consequence of interactions. For example, the characteristic 
short-range repulsion of the $NN$ interaction precludes particles from approaching 
each other. This results in a pair-correlation function that vanishes at short 
separations. In the particular case of the CIM, this short-range repulsion is enforced
by precluding two nucleons from occupying the same site. The static structure factor
is obtained from the pair-correlation function through a Fourier transform. That 
is~\cite{vesely1},
\begin{equation}
   S({\bf k}) = 1 + \frac{N}{V} \int d^{3}r \Big(g({\bf r})-1\Big)
   {\Large{e}}^{-i{\bf k}\cdot{\bf r}} \;.
 \label{SofK}
\end{equation}
Given that the static structure factor accounts for the mean-square density 
fluctuations in the ground state, it becomes a particularly useful indicator of 
the critical behavior associated with phase transitions---which themselves are 
characterized by the development of large ({\sl i.e.,} macroscopic) fluctuations. 
Indeed, the spectacular phenomenon of {\sl ``critical opalescence''} in fluids is 
the macroscopic manifestation of abnormally large density fluctuations---and 
thus abnormally large light scattering---near a phase transition~\cite{pathria1}. 
In this regard, the static structure factor at {\sl zero-momentum transfer} provides 
a unique connection to the thermodynamics of the
system~\cite{pathria1}. That is,
\begin{equation}
   S({\bf k}\!=\!0) = 
   \frac{\langle N^{2}\rangle -\langle N\rangle^{2}}{\langle N\rangle}  = 
   \frac{\langle N\rangle k_{B}T}{V}\kappa_{T} \;,
 \label{SofKZero}
\end{equation}
where $\kappa_{T}$ is the isothermal compressibility of the system. The isothermal 
compressibility is reminiscent of the heat capacity [Eq.\,(\ref{HeatCapacity})] that 
accounts for energy rather than density fluctuations. As such, they both play a critical 
role in identifying the onset of phase transitions.

As mentioned earlier, the various configurations of the system will be generated via
a Metropolis Monte-Carlo algorithm with a weighting factor determined by the total 
potential energy of the CIM [see Eq.\,(\ref{Palpha})]. The Metropolis algorithm is very 
well known\,\cite{metropolis1,vesely1}, so we only provide a brief review of those parts 
of relevance to our implementation. In particular, all Monte-Carlo moves must be 
consistent with the specified baryon number $A$ and proton fraction $x_{p}$. Thus, 
given a current configuration ${\pmb\alpha}$, we propose a move to a new 
configuration ${\pmb\alpha'}$ by selecting two lattice sites ($i$ and $j$) at random and 
then simply exchange their occupancies ({\sl i.e.,} $\alpha_{i}\!\leftrightarrow\!\alpha_{j}$). 
This move ensures that both the baryon number the proton fraction are conserved 
during the simulation. The new configuration is accepted provided 
\begin{equation}
 \frac{P_{\pmb\alpha'}(T)}{P_{\pmb\alpha}(T)} > {\rm rand} \;,
 \label{MCStep}  
\end{equation}
where ${\rm ``rand"}$ is a random number between 0 and 1 drawn from a 
uniform distribution. Otherwise, the move is rejected and the original configuration 
${\pmb\alpha}$ is kept. 

Initially, the lattice is populated by placing $Z\!=\!x_{p}A$ protons and 
$N\!=\!A\!-\!Z$ neutrons at random throughout the $S$ lattice sites. Given 
that each lattice site is occupied by at most one nucleon, a total of $S\!-\!A$ 
sites remain empty. The simulation starts by 
thermalizing the system at a temperature that is significantly higher than the 
target temperature $T$; this prescription prevents the system from getting
trapped in a local minimum. Once the system is properly thermalized at the 
higher temperature, a very slow cooling schedule is enforced until the 
desired temperature $T$ is reached. Note that without a proper cooling schedule, 
a system that should crystallize at low temperature may end up resembling an 
amorphous solid. Once the system reaches the target temperature $T$, one 
proceeds to accumulate statistics in order to compute the thermal averages 
for a variety of physical observables. However, a strong correlation is likely to
exist between two neighboring configurations ${\pmb\alpha}$ and ${\pmb\alpha}'$ 
since they differ by (at most) the permutation of two occupation numbers. This
correlations can significantly bias the results and may lead to an improper 
estimate of the Monte Carlo errors. To prevent this situation from developing, one 
selects {\sl uncorrelated} events by calculating the normalized auto-correlation 
function of a suitable observable ${\cal O}$. For a large sequence of 
configurations $\{ {\pmb\alpha}_{1}, {\pmb\alpha}_{2}, \ldots\}$, the auto-correlation
function of ${\cal O}$ is defined by the following expression:
\begin{equation}
 f\!_{{}_{\cal O}}\!(m)=\frac{\sum_{n=1}
  \Big({\cal O}_{n}\!-\!\langle{\cal O}\rangle\Big)
  \Big({\cal O}_{n+m}\!-\!\langle{\cal O}\rangle\Big)}
  {\sum_{n=1}
  \Big({\cal O}_{n}\!-\!\langle{\cal O}\rangle\Big)^{2}}\;,
  \end{equation}
where ${\cal O}_{n}\!\equiv\!{\cal O}({\pmb\alpha}_{n})$. The decorrelation {\sl ``time"}
$\tau$ is defined by the condition $f\!_{{}_{\cal O}}\!(\tau)\!=\!0.1$. 
In Fig.\,\ref{fig:autocorrelation} we display the auto-correlation function for the total 
potential energy $V_{\rm Total}$ for a filling fraction $A/S\!=\!0.2$, a proton fraction of 
$x_{p}\!=\!0.3$, and temperatures of $T\!=\!10$\,MeV and $T\!=\!15$\,MeV. At the 
lower temperature it becomes more difficult to explore the full energy landscape, 
thereby resulting a in a longer decorrelation time. For this particular case,  
$\tau_{{}_{\!10}}\!=\!7,218$ and $\tau_{{}_{\!15}}\!=\!5,550$. In what follows, all our
results are reported with a proper treatment of Monte Carlo errors.
\begin{figure}[htbp]
\begin{center}
\includegraphics[width=\columnwidth]{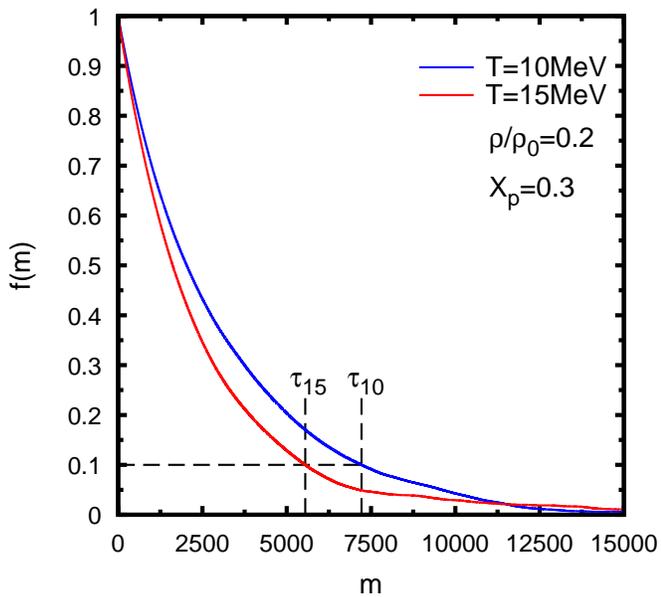}
\end{center}
\caption{The auto-correlation function for the total potential energy at fixed
density, fixed proton fraction, and two values of the temperature. The 
decorrelation time $\tau$ is defined by the condition $f(\tau)\!=\!0.1$.}
\label{fig:autocorrelation}
\end{figure}

\section{Results}
\label{Results}
We start this section by providing a baseline CIM calculation that aims to 
reproduce the results reported recently in Ref.\,\cite{piekarewicz1}. Recall
that in Ref.\,\cite{piekarewicz1} the temperature was fixed at $T\!=\!1$\,MeV
in order to simulate the quantum zero-point motion. It is the goal of our present 
lattice-gas simulation to improve on such a work by examining the role of 
the temperature on the structure and dynamics of the inner stellar crust.
Ultimately then, this sort of simulations will help us explore the phase
diagram as a function of temperature, density, and proton fraction.

Given that the static structure factor at zero momentum transfer accounts 
for density fluctuations [see Eq.\,(\ref{SofKZero})], we begin this section by
displaying in Figs.\,\ref{fig:GrPT1MeV}-\ref{fig:SqNT1MeV} pair correlations 
functions and static structure factors for neutrons and protons at a fixed 
temperature of $T\!=\!1$\,MeV. Note that because of the discrete nature of 
the lattice, all distances between sites are {\sl ``quantized"}. Moreover, due 
to the periodicity of the lattice, the allowed values of the momenta are given 
as follows:
\begin{equation}
 {\bf k}=\frac{2\pi}{L}{\bf l} \quad  (l_{i}=0,1,\ldots,S_{i}\!-\!1) \;,
\end{equation}
where $S_{x}\!=\!S_{y}\!=\!S_{z}\!=\!S^{1/3}$. 

Results are presented as a 
function of the proton fraction for a lattice of $S\!=\!(30)^{3}$ sites, and a 
filling fraction of $\rho/\rho_{{}_{\!0}}\!=\!A/S\!=\!0.1875$ (or 
$A\!\simeq\!5,000$ nucleons). The pair correlation function is characterized 
by a set of discrete peaks at the allowed distances on the lattice. For example,
at this relatively low filling fraction, the dynamics favors the formation of 
neutron-rich clusters immersed in a dilute neutron vapor (see 
Fig.\,\ref{fig:snapshot1MeV}). Given that the neutron-proton interaction is
attractive, nucleons organize themselves within a cluster by occupying 
alternating lattice sites. Thus, the closest distance between nucleons of 
the same species is $r_{\rm min}\!=\!\sqrt{2}a\!=\!2.605\,{\rm fm}$, where
$a\!=1.842\,{\rm fm}$ is the lattice spacing [see Eq.\,(\ref{aLatt})]. The 
largest peak in both Figs.\,\ref{fig:GrPT1MeV} and\,\ref{fig:GrNT1MeV} 
reflect this behavior. In the case of protons---where no dilute vapor is
formed---other peaks corresponding to more distant protons are clearly 
discernible at distances of $2a\!=\!3.684$, $\sqrt{6}a\!=\!4.512$, 
$\sqrt{8}a\!=\!5.210$, $\sqrt{10}a\!=\!5.825,\ldots$ In the case of neutrons, 
the existence of a dilute neutron vapor gives rise to additional peaks and 
to significant pair-correlation strength at larger distances. 
\begin{figure}[htbp]
\begin{center}
\includegraphics[width=\columnwidth]{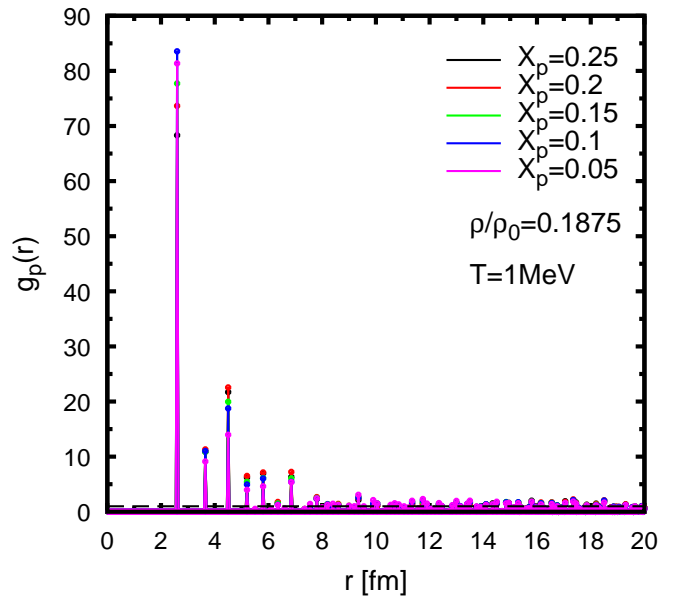}
\end{center}
\caption{Proton pair correlation function for different proton fractions 
at a fixed density of $\rho/\rho_{{}_{0}}\!=\!0.1875$ and a temperature 
of $T\!=\!1$\,MeV.} 
\label{fig:GrPT1MeV}
\end{figure}
\begin{figure}[htbp]
\begin{center}
\includegraphics[width=\columnwidth]{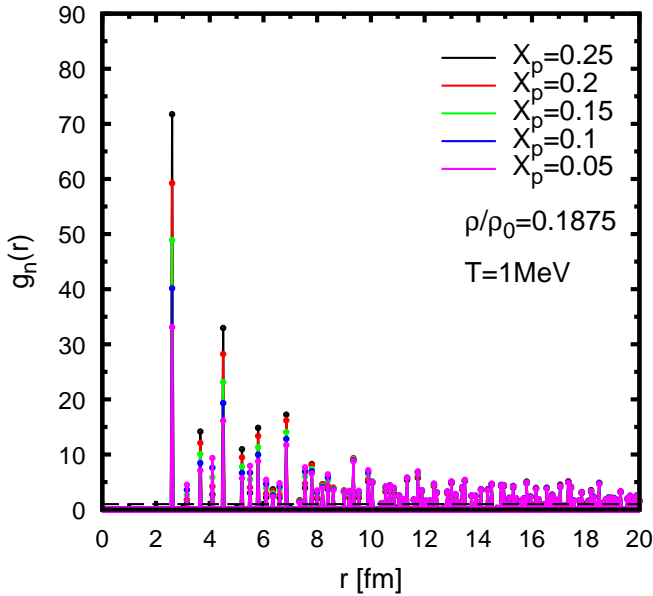}
\end{center}
\caption{Neutron pair correlation function for different proton fractions 
at a fixed density of $\rho/\rho_{{}_{0}}\!=\!0.1875$ and a temperature 
of $T\!=\!1$\,MeV.}
\label{fig:GrNT1MeV}
\end{figure}
The corresponding static structure factors for both protons and neutrons are shown in
Figs.\,\ref{fig:SqPT1MeV} and \ref{fig:SqNT1MeV}, respectively. Our results reproduce
qualitatively those reported in Ref.\,\cite{piekarewicz1}. That is, $S(k)$ displays a 
prominent peak that becomes progressively higher with increasing proton fraction. 
The peak in $S(k)$ occurs at a momentum transfer $k$ for which the probe 
({\sl e.g.,} electrons in the case of protons and neutrinos in the case of neutrons) 
can most efficiently scatter from the density fluctuations in the system. In particular, 
if the wavelength of the probe is large as compared with the size of the pasta structures, 
the scattering may be coherent. This can significantly enhance the response or,
equivalently, significantly reduce the electron/neutrino mean-free path. Finally, as
in Ref.\,\cite{piekarewicz1}, there is no visible enhancement in $S(k)$ at zero 
momentum transfer as would be expected from the putative phase transition from 
a Wigner Crystal to a pasta phase.
\begin{figure}[htbp]
\begin{center}
\includegraphics[width=\columnwidth]{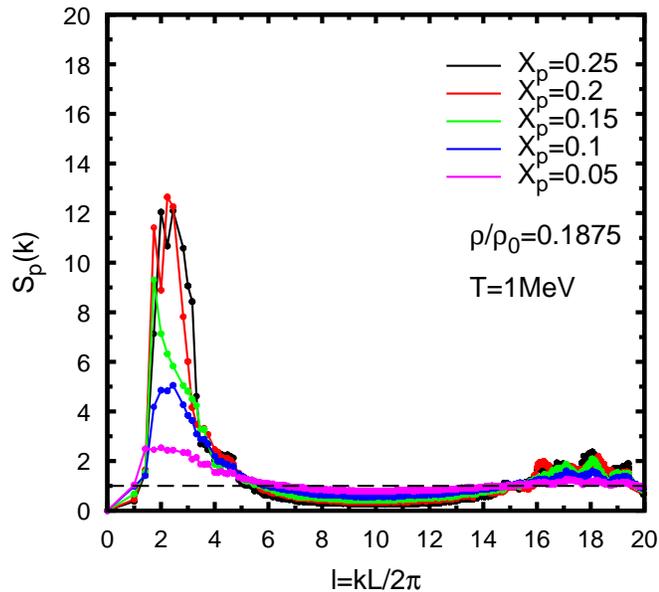}
\end{center}
\caption{Proton static structure factor for different proton fractions 
at a fixed density of $\rho/\rho_{{}_{0}}\!=\!0.1875$ and a
temperature of $T\!=\!1$\,MeV.}
\label{fig:SqPT1MeV}
\end{figure}
\begin{figure}[htbp]
\begin{center}
\includegraphics[width=\columnwidth]{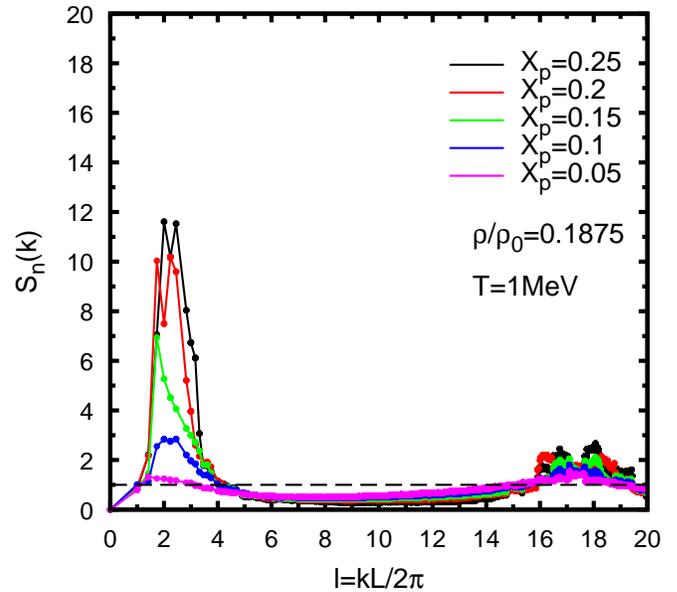}
\end{center}
\caption{Neutron static structure factor for different proton fractions 
at a fixed density of $\rho/\rho_{{}_{0}}\!=\!0.1875$ and a
temperature of $T\!=\!1$\,MeV.}
\label{fig:SqNT1MeV}
\end{figure}
Although it is gratifying that the CIM reproduces the trends reported in 
Ref.\,\cite{piekarewicz1}, an important goal of the present work is to 
explore the evolution of the system---particularly the dissolution of the 
pasta---as a function of temperature. In analogy to the static structure
factor that captures the density fluctuations in the system, we now
examine thermal fluctuations through a study of the heat 
capacity [see Eq.\,(\ref{HeatCapacity})]. We start by displaying in
Figs.\,\ref{fig:GrPXp3}-\ref{fig:SqNXp3} the neutron and proton pair 
correlation functions and corresponding static structure factors at a 
fixed density of $\rho/\rho_{{}_{0}}\!=\!0.2$, a fixed proton fraction of 
$x_{p}\!=\!0.3$, and for temperatures ranging from $T\!=\!1\,{\rm MeV}$ 
to $T\!=\!6\,{\rm MeV}$. In particular, note that under these conditions 
of density and proton fraction---and at the low temperature of 
$T\!=\!1\,{\rm MeV}$---the existence of a pasta phase has been well 
established\,\cite{horowitz1,watanabe2}; see also Fig.\,\ref{fig:snapshot1MeV}.
\begin{figure}[htbp]
\begin{center}
\includegraphics[width=\columnwidth]{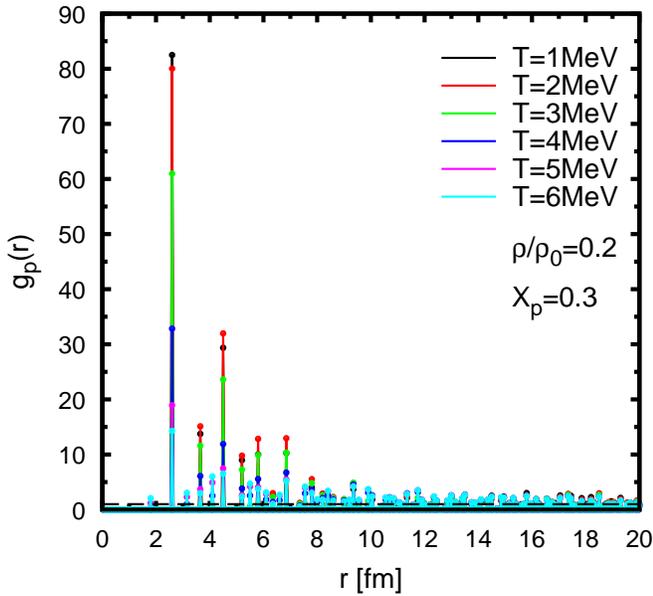}
\end{center}
\caption{Proton pair correlation function for various temperatures,
at a fixed density of $\rho/\rho_{{}_{0}}\!=\!0.2$, and a proton 
fraction of $x_{p}\!=\!0.3$.}
\label{fig:GrPXp3}
\end{figure}
\begin{figure}[htbp]
\begin{center}
\includegraphics[width=\columnwidth]{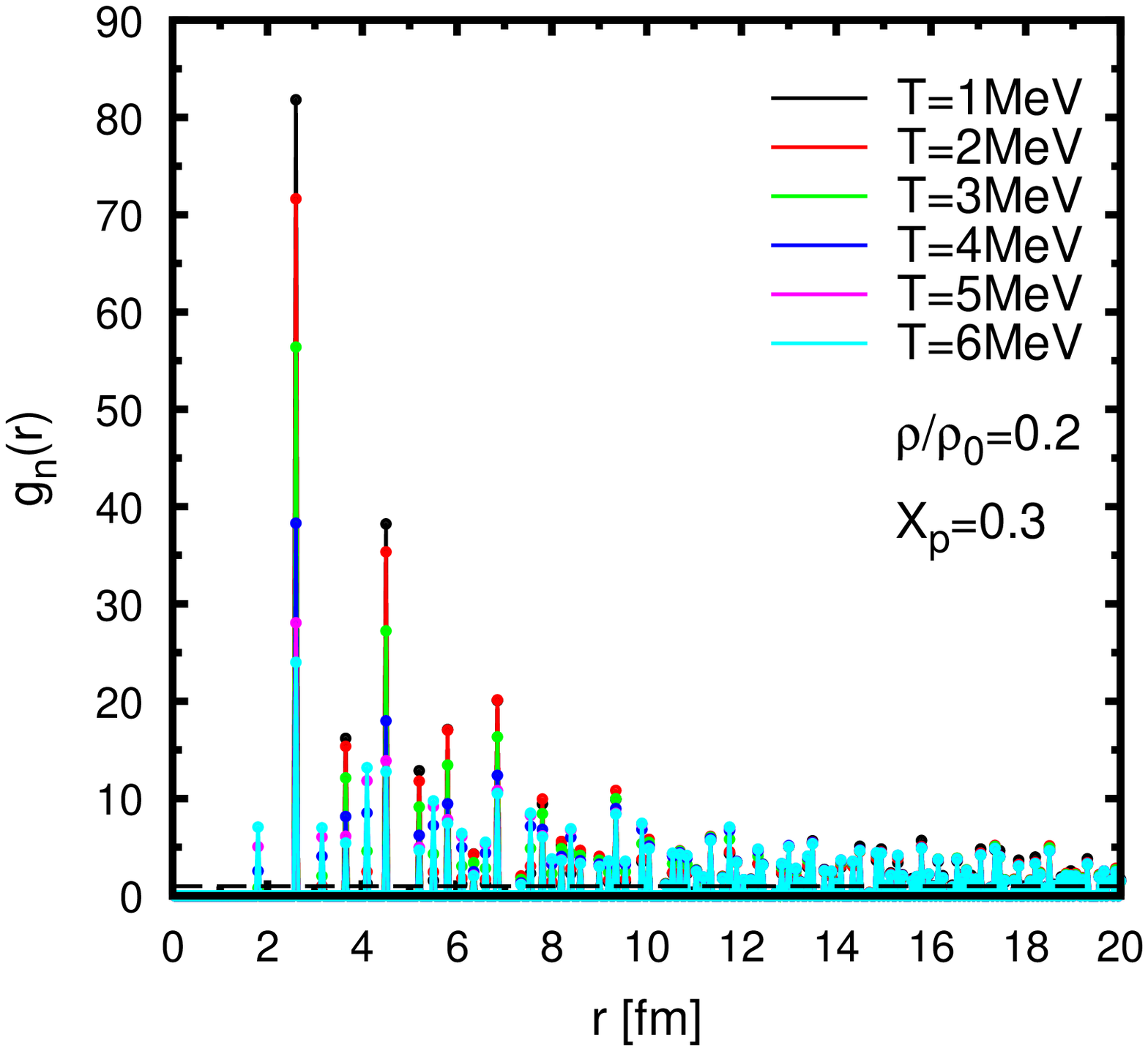}
\end{center}
\caption{Neutron pair correlation function for various temperatures,
at a fixed density of $\rho/\rho_{{}_{0}}\!=\!0.2$, and a proton 
fraction of $x_{p}\!=\!0.3$.}
\label{fig:GrNXp3}
\end{figure}
\begin{figure}[htbp]
\begin{center}
\includegraphics[width=\columnwidth]{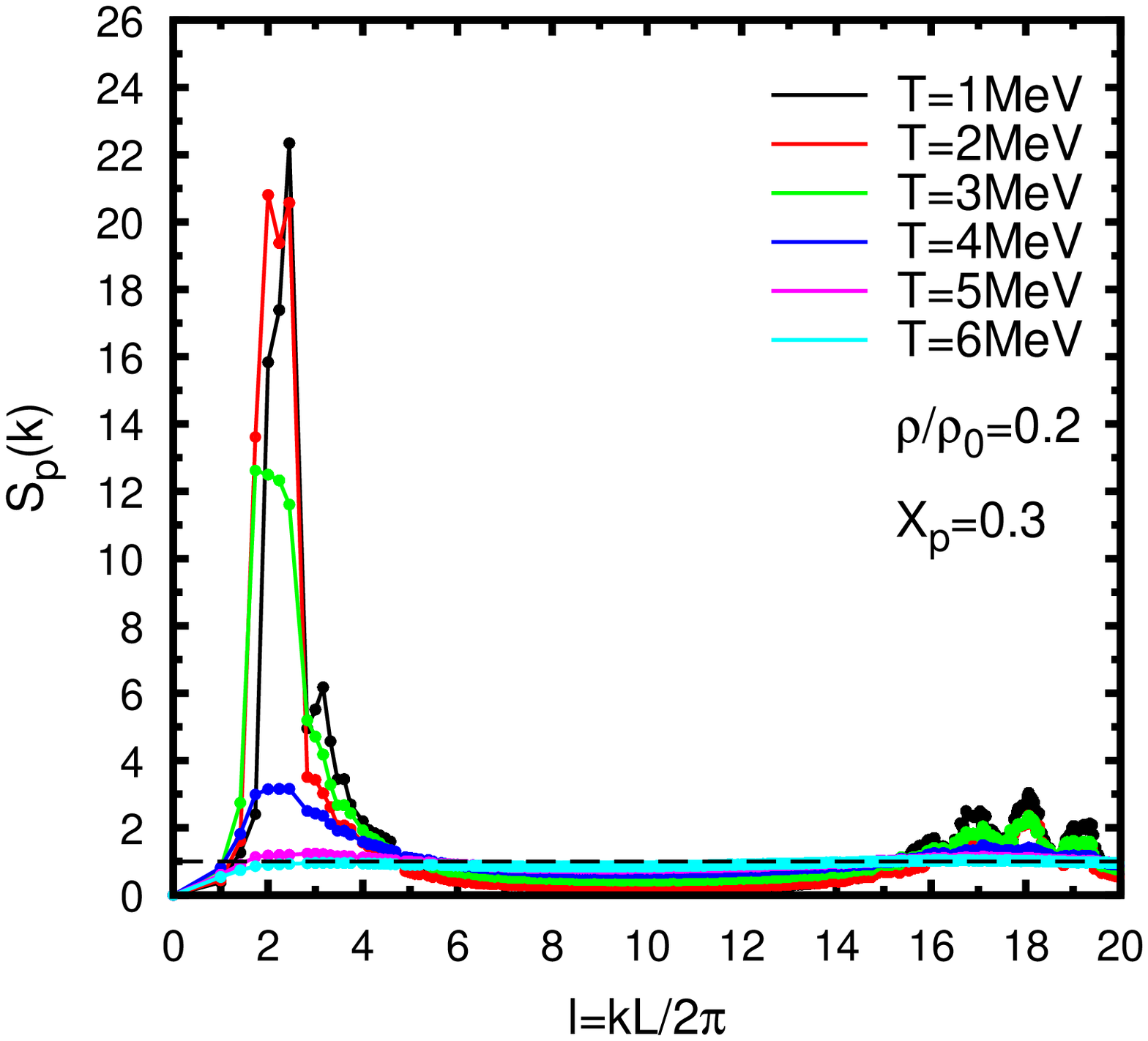}
\end{center}
\caption{Proton static structure form factor for various temperatures,
at a fixed density of $\rho\!/\!\rho_{{}_{0}}\!=\!0.2$, and a proton 
fraction of $x_{p}\!=\!0.3$.}
\label{fig:SqPXp3}
\end{figure}
\begin{figure}[htbp]
\begin{center}
\includegraphics[width=\columnwidth]{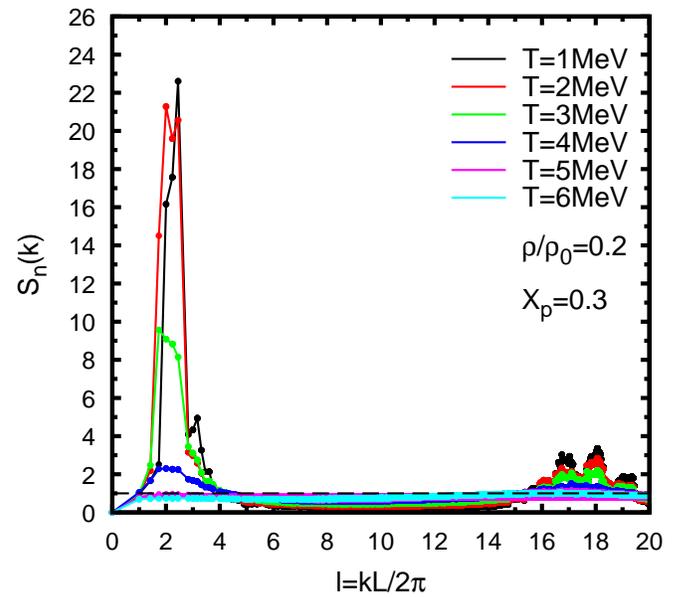}
\end{center}
\caption{Neutron static structure form factor for various temperatures,
at a fixed density of $\rho/\rho_{{}_{0}}\!=\!0.2$, and a proton 
fraction of $x_{p}\!=\!0.3$.}
\label{fig:SqNXp3}
\end{figure}

The $T\!=\!1\,{\rm MeV}$ results set the baseline as these can be directly
compared against our earlier findings. As before, at low temperatures 
($T\!\lesssim\!3\,{\rm MeV}$) the system displays the strong clustering 
correlations characteristic of the pasta phase. However, as the temperature
increases and the thermal energy becomes comparable to the binding 
energy per nucleon of the neutron-rich clusters the behavior changes 
dramatically. The large peaks in both the pair correlation function and
the static structure factor get significantly reduced as the system reaches
a temperature of $T\!\simeq\!4\,{\rm MeV}$ and both become essentially 
structurless at $T\!\gtrsim\!5\,{\rm MeV}$. Note also the appearance of
a small peak in $g_{n}(r)$ at $r\!=\!a$ as the entropic contribution starts 
to become as important, if not more, than the energy contribution.
\begin{figure}[htbp]
\begin{center}
\includegraphics[width=\columnwidth]{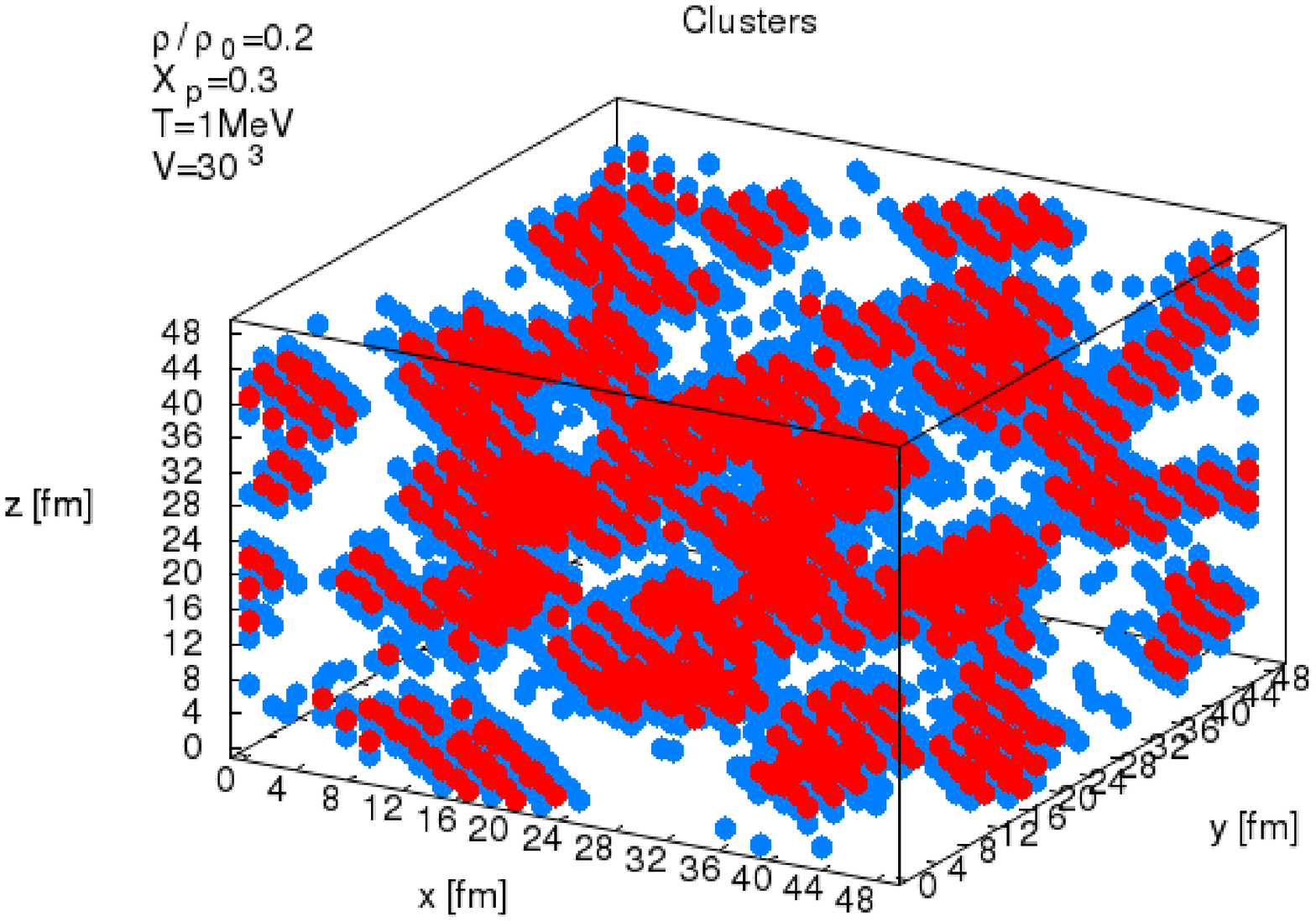}
\end{center}
\caption{Monte-Carlo snapshot of neutron-star matter at a temperature
of $T\!=\!1\,{\rm MeV}$, a density of $\rho/\rho_{{}_{0}}\!=\!0.2$, and 
a proton fraction of $x_{p}\!=\!0.3$. The blue and red dots are used to
display the location of neutrons and protons, respectively.}
\label{fig:snapshot1MeV}
\end{figure}
\begin{figure}[htbp]
\begin{center}
\includegraphics[width=\columnwidth]{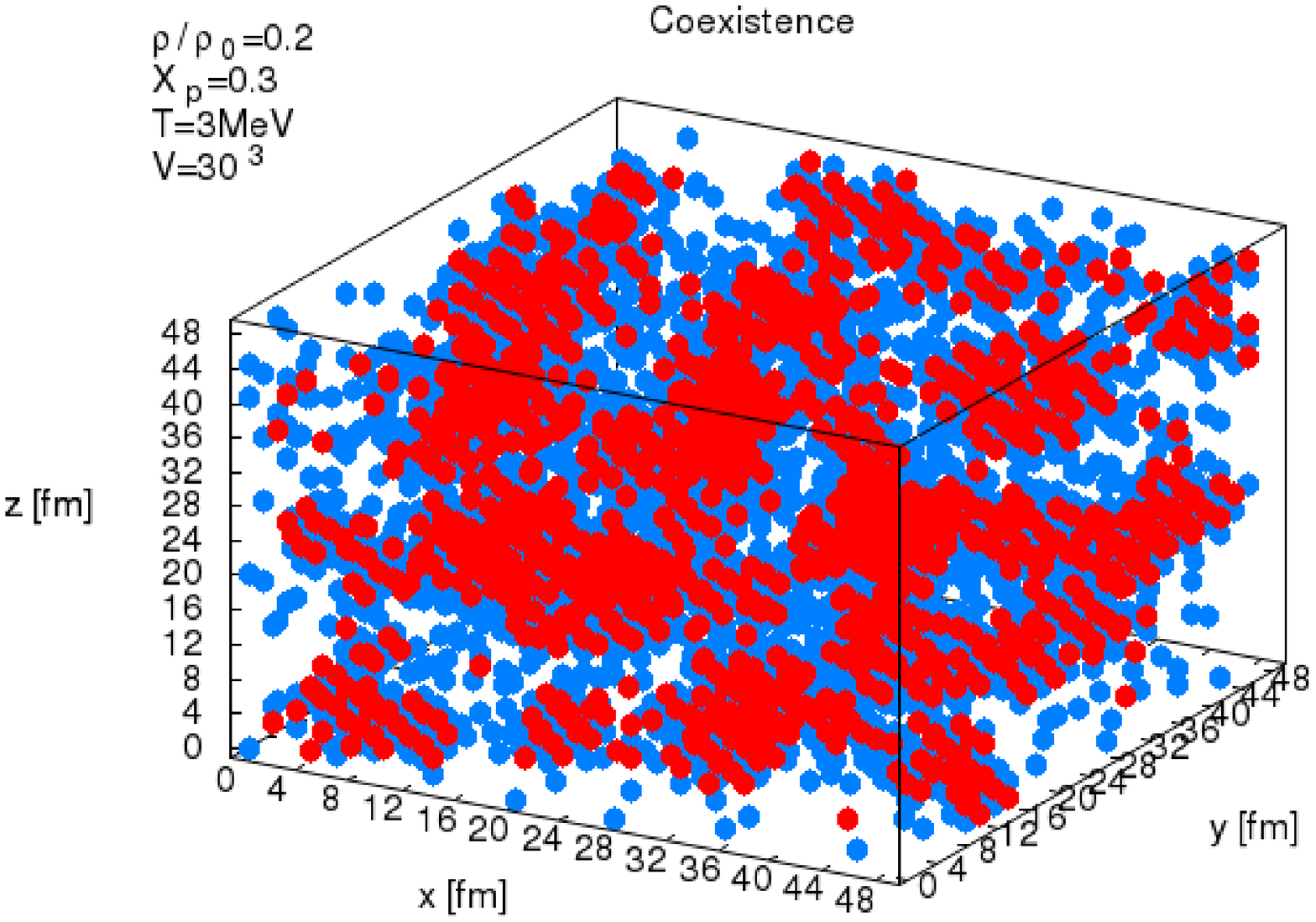}
\end{center}
\caption{Monte-Carlo snapshot of neutron-star matter at a temperature 
of $T\!=\!3\,{\rm MeV}$, a density of $\rho/\rho_{{}_{0}}\!=\!0.2$, and a 
proton fraction of $x_{p}\!=\!0.3$. The blue and red dots are used to
display the location of neutrons and protons, respectively.}
\label{fig:snapshot4MeV}
\end{figure}
\begin{figure}[htbp]
\begin{center}
\includegraphics[width=\columnwidth]{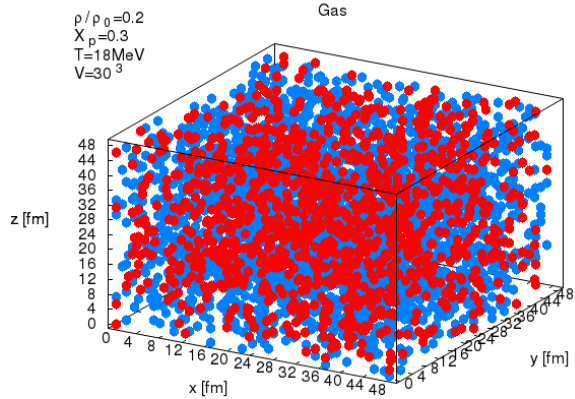}
\end{center}
\caption{Monte-Carlo snapshot of neutron-star matter at a temperature 
of $T\!=\!18\,{\rm MeV}$, a density of $\rho/\rho_{{}_{0}}\!=\!0.2$, and a 
proton fraction of $x_{p}\!=\!0.3$. The blue and red dots are used to
display the location of neutrons and protons, respectively.}
\label{fig:snapshot18MeV}
\end{figure}
To illustrate the behavior of the system as a function of temperature we 
display in Figs.\,\ref{fig:snapshot1MeV}-\ref{fig:snapshot18MeV} Monte-Carlo 
snapshots at a density of $\rho/\rho_{{}_{0}}\!=\!0.2$, a proton fraction of 
$x_{p}\!=\!0.3$, and temperatures of $T\!=\!1\,{\rm MeV}$, $T\!=\!3\,{\rm MeV}$, 
and $T\!=\!18\,{\rm MeV}$. One can see the gradual transition in the structure of
the system. At $T\!=\!1\,{\rm MeV}$ the system displays the existence of 
neutron-rich clusters surrounded by a dilute neutron vapor. As the temperature
is increased to $T\!=\!3\,{\rm MeV}$, some of the weakly bound neutrons in the
clusters join the vapor and one sees a coexistence between the clusters and
the vapor. Finally, at the very large temperature of $T\!=\!18\,{\rm MeV}$ no
spatial correlations remain as the system has been fully vaporized into a 
classical gas of nucleons. Quantitatively, this behavior is captured by the
heat capacity which has been computed by calculating the fluctuations in
energy [see Eq.\,(\ref{HeatCapacity})] and is displayed in Fig.\,\ref{fig:Cv}. 
The energy fluctuations are small in both the clustered and gas phases---but 
increases significantly at $T\!\simeq\!3\,{\rm MeV}$ where both phases coexists.
\begin{figure}[htbp]
\begin{center}
\includegraphics[width=\columnwidth]{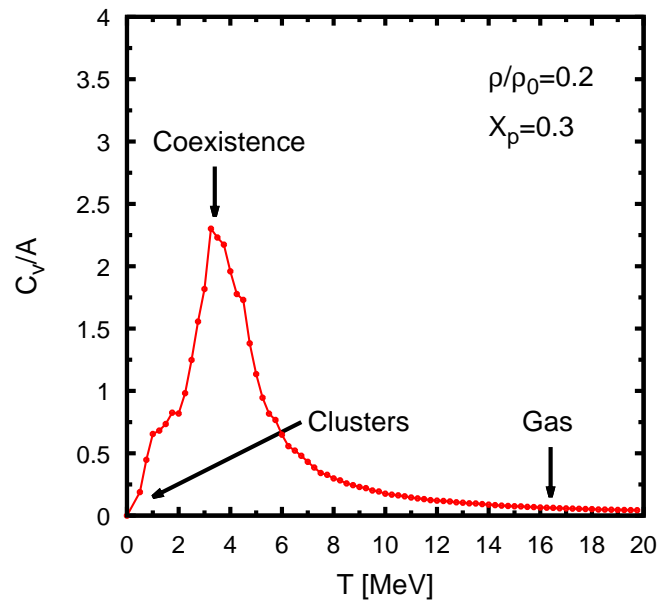}
\end{center}
\caption{The heat capacity of neutron-star matter as a function of temperature 
and at a density of $\rho/\rho_{{}_{0}}\!=\!0.2$ and a proton fraction of 
$x_{p}\!=\!0.3$.}
\label{fig:Cv}
\end{figure}

\section{Conclusions and Outlook}
\label{Conclusions}
The quest for physical observables that are particularly sensitive to pasta formation
remains elusive. Indeed, even the existence of the pasta phase in the proton-deficient 
environment of the inner stellar crust remains an open question. In the present
work we introduced a new model---the \sl ``Charged Ising Model''}---to tackle some
of these fundamental questions. The CIM is a lattice-gas model that while simple 
in its assumptions, retains the essence of Coulomb frustration that is required to 
capture the subtle dynamics of the inner stellar crust. Monte Carlo simulations 
based on this model are not as computationally demanding because the long-range 
Coulomb interaction---computed here exactly via an Ewald summation---was 
pre-computed at the appropriate lattice sites and then stored in memory prior to the 
start of the simulation. This represents an enormous advantage in simulating systems 
with a large number of particles as a function of density, temperature, and proton 
fraction. In this first---and mostly exploratory study---we were able to simulate 
systems with as many as $S\!=\!30^3$ lattice sites as a function of temperature,
density, and proton fraction. Particular attention was placed on physical observables
such as the pair correlation function, the corresponding static structure factor, and
the heat capacity; quantities that properly capture both density and thermal fluctuations
in the system. Note that the static structure factor displays a prominent coherent peak 
that occurs at a momentum transfer for which the probe ({\sl e.g.,} electrons in the case 
of protons and neutrinos in the case of neutrons) can most efficiently scatter from the 
density fluctuations in the system. The existence of pasta structures can therefore 
significantly reduce the electron or neutrino mean-free path in the stellar crust. A 
detailed study of electron and neutrino transport within the CIM will be forthcoming.
We note that the very simple nuclear part of the Hamiltonian 
[Eqs.\,(\ref{VNuclear}) and\,(\ref{VNTwoBody})] essentially represents an Ising 
Hamiltonian for a spin-one system. That is,
\begin{equation}
 V_{\rm Nuclear} = \frac{\epsilon}{2}\sum_{\langle i,j\rangle}^{S}n_{i}n_{j} \,, 
 \label{VNuclear1}
\end{equation}
where $\epsilon\!\simeq\!16/3\,{\rm MeV}$ but now $n_{i}\!=\!-1,0,1$ can take 
three values depending on whether the site is occupied by a neutron, is empty, 
or is occupied by a proton. To properly simulate neutron-star matter, this
Hamiltonian---together with the long-range Coulomb part---must be solved at
constant magnetization rather than at constant magnetic field. Although much
work has been done along the lines of the Ising model, the virtue of the CIM
is that it incorporates the relatively simple spin-1 Ising model together with 
the challenging long-range Coulomb interactions.

Finally, in the future we plan to use the CIM to carry out an analysis similar 
to the one recently reported in Ref.\,\cite{Pramudya2011}. In such a 
condensed-matter study a large temperature gap, {\sl i.e.,}  
$k_{B}T_{c}\!<\!k_{B}T\!\ll\!E_{\rm Coul}$, was identified between the 
melting temperature and the Coulomb energy where the system displays
unconventional ``pasta-like'' behavior as a result of the strong frustration 
induced by the long-range interactions. Particularly relevant is the emergence 
of a {\sl pseudogap} in the density of states that appears to mediate the transition 
from the Wigner Crystal to the uniform Fermi liquid. We are confident that the 
evolution of the pseudogap region as a function of proton fraction may help us
prove the existence---or lack-thereof---of a pasta phase in the inner stellar
crust. 

\medskip
\begin{acknowledgments}
 We thank Prof. Vladimir Dobrosavljevi\'c and Yohanes Pramudya
 for many fruitful discussions. This work was supported in part by 
 grant DE-FD05-92ER40750 from the U.S. Department of Energy 
 and partly by the ANR under the project NExEN ANR-07-BLAN-0256-02.
\end{acknowledgments}


\bibliography{../ReferencesJP}

\begin{thebibliography}{10}

\bibitem{oppenheimer1}
J. R. Oppenheimer and G. M. Volkoff,
\newblock Phys. Rev. \textbf{55}, 374 (1939).

\bibitem{bombaci1}
I. Bombaci,
\newblock  A\&A \textbf{305}, 871 (1996).

\bibitem{glendenning1}
N. K. Glendenning,
\newblock Compact Stars: Nuclear Physics, Particle Physics, and 
General Relativity, Springer-Verlag (2000).

\bibitem{glendenning2}
N. K. Glendenning,
\newblock  Phys. Rep. \textbf{342}, 393 (2001).

\bibitem{NucPhys2012}
\newblock The Committee on the Assessment of and Outlook for Nuclear Physics; 
 Board on Physics and Astronomy; Division on Engineering and Physical Sciences; 
 National Research Council, {\sl ``Nuclear Physics: Exploring the Heart of Matter"}, 
 The National Academies Press (2012).


\bibitem{Ellis:1995kz}
P. J. Ellis, R. Knorren, and M. Prakash,
\newblock Phys. Lett. B {\bf 349}, 11 (1995).

\bibitem{Pons:2000xf}
J. A. Pons, J. A. Miralles, M. Prakash, and J. M. Lattimer,
\newblock Astrophys. J. {\bf 553}, 382 (2001).

\bibitem{Weber:2004kj}
F. Weber, 
\newblock Prog. Part. Nucl. Phys. {\bf 54}, 193 (2005).

\bibitem{Alford:1998mk}
M. G. Alford, K. Rajagopal, and F. Wilczek,
\newblock Nucl. Phys. B {\bf 537}, 443 (1999).

\bibitem{Alford:2007xm}
M. G. Alford, A. Schmitt, K. Rajagopal, and T. Schafer,
\newblock Rev. Mod. Phys. {\bf 80}, 1455 (2008).

\bibitem{horowitz1}
C. J. Horowitz, M. A. P\'erez-Garc\'ia, and J. Piekarewicz,
\newblock Phys. Rev. C {\bf 69}, 045804 (2004).

\bibitem{watanabe1}
G. Watanabe, T. Maruyama, K. Sato, K. Yasuoka, and T. Ebisuzaki,
\newblock Phys. Rev. Lett. {\bf 94}, 031101 (2005).

\bibitem{margueron1}
J\'er\^ome Margueron, Jes\'us Navarro, and Patrick Blottiau,
\newblock Phys. Rev. C {\bf 70}, 028801 (2004).

\bibitem{chamel1}
N. Chamel and P. Haensel,
\newblock Living Rev. Relativity \textbf{11} (2008), 10.

\bibitem{bertulani1}
C. Bertulani and J. Piekarewicz editors,
\newblock  {\sl ``Neutron Star Crust"}, Nova Publishers (2012).

\bibitem{ravenhall1}
D. G. Ravenhall, C. J. Pethick, and J. R. Wilson,
\newblock Phys. Rev. Lett. {\bf 50}, 2066 (1983).

\bibitem{lorentz1}
C. P. Lorenz, D. G. Ravenhall, and C. J. Pethick,
\newblock Phys. Rev. Lett. {\bf 70}, 379 (1993). 

\bibitem{ravenhall2}
D.G. Ravenhall and C.J. Pethick,
\newblock Annu. Rev. Nucl. Part. Sci. {\bf 45} 429 (1995).

\bibitem{horowitz2}
C. J. Horowitz, M. A. P\'erez-Garc\'ia, J. Carriere, D. K. Berry, and J. Piekarewicz,
\newblock Phys. Rev. C {\bf 70}, 065806 (2004).

\bibitem{horowitz3}
C. J. Horowitz, M. A. P\'erez-Garc\'ia, D. K. Berry, and J. Piekarewicz,
\newblock Phys. Rev. C {\bf 72}, 035801 (2005).

\bibitem{maruyama1}
T. Maruyama, K. Niita, K. Oyamatsu, T. Maruyama, S. Chiba, and A. Iwamoto,
\newblock Phys. Rev. C {\bf 57}, 655 (1998). 

\bibitem{watanabe2}
G. Watanabe, K. Sato, K. Yasuoka, and T. Ebisuzaki,
\newblock Phys. Rev. C {\bf 66}, 012801 (2002).

\bibitem{watanabe3}
G. Watanabe, K. Sato, K. Yasuoka, and T. Ebisuzaki,
\newblock Phys. Rev. C {\bf 68}, 035806 (2003).

\bibitem{watanabe4}
G. Watanabe, K. Sato, K. Yasuoka, and T. Ebisuzaki,
\newblock Phys. Rev. C {\bf 69}, 055805 (2004).

\bibitem{watanabe5}
H. Sonoda, G. Watanabe, K. Sato, T. Takiwaki, K. Yasuoka, and T. Ebisuzaki,
\newblock Phys. Rev. C {\bf 75}, 042801 (2007).

\bibitem{watanabe6}
H. Sonoda, G. Watanabe, K. Sato, K. Yasuoka, and T. Ebisuzaki,
\newblock Phys. Rev. C {\bf 77}, 035806 (2008).

\bibitem{watanabe7}
G. Watanabe, H. Sonoda, T. Maruyama, K. Sato, K. Yasuoka, and T. Ebisuzaki,
\newblock Phys. Rev. Lett. {\bf 103}, 121101 (2009).

\bibitem{watanabe8}
G. Watanabe, K. Sato, K. Yasuoka, and T. Ebisuzaki,
\newblock Phys. Rev. C {\bf 81}, 049901 (2010).

\bibitem{watanabe9}
H. Sonoda, G. Watanabe, K. Sato, K. Yasuoka, and T. Ebisuzaki,
\newblock Phys. Rev. C {\bf 81}, 049902 (2010).

\bibitem{sebille1}
B. Jouault, F. S\'ebille, and V. de la Mota,
\newblock Nucl. Phy. {\bf A628}, 119 (1998).

\bibitem{sebille2}
F. S\'ebille, V. de la Mota, and S. Figerou,
\newblock Nucl. Phy. {\bf A822}, 51 (2009).

\bibitem{sebille3}
F. S\'ebille, V. de la Mota, and S. Figerou,
\newblock Phys. Rev. C {\bf 84}, 055801 (2011).

\bibitem{maruyama2}
T. Maruyama, T. Tatsumi, D. N. Voskresensky, T. Tanigawa, and S. Chiba
\newblock Phys. Rev. C {\bf 72}, 015802 (2005).

\bibitem{maruyama3}
T. Maruyama, T. Tatsumi, D. N. Voskresensky, T. Tanigawa, T. Endo, and S. Chiba
\newblock Phys. Rev. C {\bf 73}, 035802 (2006).

\bibitem{avancini1}
S. S. Avancini, D. P. Menezes, M. D. Alloy, J. R. Marinelli, M. M. W. Moraes, and C. Provid\^encia,
\newblock Phys. Rev. C {\bf 78}, 015802 (2008). 

\bibitem{avancini2}
S. S. Avancini, L. Brito, J. R. Marinelli, D. P. Menezes, M. M. W. de Moraes, C. Providência, and A. M. Santos
\newblock Phys. Rev. C 79, 035804 (2009).

\bibitem{grill1}
F. Grill, C. Provid\^encia, and S. S. Avancini,
\newblock Phys. Rev. C {\bf 85}, 055808 (2012).

\bibitem{newton}
W. G. Newton and J. R. Stone,
\newblock Phys. Rev. C {\bf 79}, 055801 (2009).

\bibitem{piekarewicz1}
J. Piekarewicz and G. Toledo S\'anchez,
\newblock Phys. Rev. C {\bf 85}, 015807 (2012).

\bibitem{Pramudya2011}
Y. Pramudya, H. Terletska, S. Pankov, E. Manousakis, and V. Dobrosavljevi\ifmmode \acute{c}\else \'{c}\fi{},
\newblock Phys. Rev. B {\bf 84}, 125120 (2011).

\bibitem{ising_star1}
P. Napolitani, Ph. Chomaz, F. Gulminelli, and K. H. O. Hasnaoui,
\newblock Phys. Rev. Lett. {\bf 98}, 131102 (2007).

\bibitem{ising_star2}
C. Ducoin, K. H. O. Hasnaoui, P. Napolitani, Ph. Chomaz, and F. Gulminelli,
\newblock Phys. Rev. C {\bf 75}, 065805 (2007).

\bibitem{ising_star3}
Ph. Chomaz, C. Ducoin, F. Gulminelli, K. Hasnaoui, and P. Napolitan,
\newblock Nucl. Phy. {\bf A787}, 603 (2007).

\bibitem{ewald1}
P. P. Ewald,
\newblock Ann. Phys. {\bf 369}, 253 (1921).

\bibitem{ewald2}
M. P. Allen, D. J . Tildesley,
\newblock Computer Simulation of liquids, Clarendon, London (1987).

\bibitem{ewald3}
A. Y. Abdulnour and J.A. Board Jr,
\newblock Computer Physics Communications {\bf 95} 73 (1996).

\bibitem{metropolis1}
N. Metropolis, A. W. Rosenbluth, M. N. Rosenbluth, A. H. Teller, and E. Teller,
\newblock J. Chem. Phys. {\bf 21}, 1087 (1953).

\bibitem{fetter1}
A. L. Fetter and J. D. Walecka,
\newblock Quantum Theory of Many Particle Systems, McGraw-Hill, New York (1971).

\bibitem{vesely1}
F. J. Vesely, 
\newblock Computational Physics: An Introduction, Kluwer Academic, New York (2001).
 
\bibitem{pathria1}
R. K. Pathria, 
\newblock Statistical Mechanics, Butterworth-Heinemann, Oxford (1996).  
 
\end{thebibliography}

\end{document}